\documentclass[a4paper,12pt]{article}
\usepackage{amsmath,amsfonts,amssymb,epsfig,subfigure,color}

\newcommand{\be}{\begin{equation}}
\newcommand{\en}{\end{equation}}

\newcommand{\filler}{\hspace*{\fill}}

\renewcommand{\vec}[1]{\boldsymbol{#1}}
\begin{document}
\numberwithin{equation}{section}

%+++++++++++++++++++++++++++++++

\title{On the Nonlinear Theory of Viscoelasticity of Differential Type}

%+++++++++++++++++++++++++++++++
\author{Edvige Pucci,
Giuseppe Saccomandi, \filler \\
{\it\small Dipartimento di Ingegneria Industriale,}
\filler\\[-6pt]
{\it\small Universit\`{a} degli Studi di Perugia,
  06125 Perugia, Italy.} \filler}

%++++++++++++++++++++++++++++++++++++++++++++++++++++++++++
\maketitle
%++++++++++++++++++++++++++++++++++++++++++++++++++++++++++
\begin{abstract}
 We consider nonlinear viscoelastic materials of differential type and for some special models we derive exact solutions of initial boundary value problems. These exact solutions are used to investigate the reasons of non-existence of global solutions for such equations. 
%%%%%%%%%%%%%%%%%%%%%%

\end{abstract}

%%%%%%%%%%%%%%%%%%%%%%

\section{Introduction and basic equations}

%%%%%%%%%%%%%%%%%%%%%%
In empirical rheology the standard sketch of the one dimensional KelvinÐVoigt model of linear viscoelasticity is of a purely viscous damper and a purely elastic spring connected in parallel. In continuum mechanics, this model may be recast in the class of the simple materials denoted as viscoelastic solids of differential type. For such materials, in the isotropic case, the Cauchy stress tensor, $\vec{T}$, is defined by the constitutive equation
\be \label{c1}
\vec{T}=\vec{T}(\vec{B};\vec{D}),
\en
where $\vec{B}=\vec{F}\vec{F}^{T}$ and $\vec{D}=(\dot{\vec{F}}\vec{F}^{-1}+\vec{F}^{-T}\dot{\vec{F}}^{T})/2$ are the left Cauchy-Green strain tensor and the stretching tensor respectively (Truesdell and Noll, 1965). (Here $\vec{F}$ is the gradient of deformation). Therefore $\vec{B}$ stands for the elastic spring and $\vec{D}$ stands for the viscous damper.  

Nowadays there is an increasing interest in the study of nonlinear models of viscoelasticity and this because in the experimental studies of rubber-like materials and soft tissues the viscous and nonlinear effects are prominent and clearly measurable (Lakes, 2009). For these reasons the possibility to elaborate feasible and reliable mathematical models of nonlinear viscoelasticity is a major problem of contemporary continuum mechanics (Wineman, 2009).  In this framework the class of constitutive equations \eqref{c1} seems to have a central role in view of its direct and clear connection with the linear Kelvin-Voigt model. 

The aim of the present note is to show that many of the nonlinear models contained  in the constitutive equation  \eqref{c1} are not \emph{acceptable} because they are clearly unphysical. 
 
To make the algebra simple as possible we shall restrict our attention to the following family of shearing motions 
\be
x =X+u(Z,t), \qquad y=Y, \qquad z =Z.
\label{01}
\en
Moreover, it is convenient to consider only incompressible materials.  In the case of compressible materials some complications arise, because the family of motions \eqref{01}  is in general not admissible (Destrade and Saccomandi, 2005) and because the constitutive issues are more complex.

In this framework, it is straightforward to check that the balance equations (in the absence of body forces) reduce to the scalar equations 
\be \label{01b}
\rho u_{tt}=\partial T_{13}/\partial Z+p_x, \quad p_y=0, \quad p_z=0,
\en
where $\rho$ is the (constant) density of the material and $p$ is the Lagrange multiplier associated with the constraint of incompressibility ($det \vec{F}=1$).  For an isotropic material the classical representation formula (Truesdell and Noll, 1965) gives
\be \label{c2}
T_{13}=Q(K^2)+\nu(K^2, K_t^2, KK_t)K_t
\en
where $K(Z,t)=u_Z$ is the amount of shear, $Q=Q(K^2)$ is the generalized shear modulus and $\nu=\nu(K^2, K_t^2, KK_t)$ is the viscosity \emph{function}.
The unknowns of our problem are the pressure field $p=p(x,y,z,t)$ and the shearing motion $u=u(Z,t)$ (or equivalently the amount of shear $K$). 

From the three scalar equations \eqref{01b} we first deduce that 
$$p(x,t)=f(t)x+g(t),$$
where $g(t)$ is an unessential function. Then by derivation, with respect of $Z$, of the first equation in \eqref{01b} we obtain 
\be   
\rho K_{tt}=\frac{\partial }{\partial Z^2} \left[Q(K^2)K+\nu(K^2, K_t^2, KK_t)K_t\right] \label{02}.
\en 
Once that \eqref{02} has been solved for $K(Z,t)$ and we integrate to obtain $u(Z,t)$ from \eqref{01b} it is possible to recover the pressure field, i.e. the function $f(t)$.    

Assuming that there is a characteristic time, $\omega^{-1}$, such that $\tau=\omega t$,  a characteristic length, $L$ ,such that $\zeta=Z/L$ and assuming that the limit
$$\mu_0=lim_{K \rightarrow 0} Q$$ 
is a positive number, we define the quantities
$$
\epsilon=\frac{\omega^2 L^2\rho}{\mu_0}, \quad \hat{Q}=\frac{Q}{\mu_0}, \quad \hat{\nu}=\frac{\omega \nu}{\mu_0},
$$ 
such that the dimensionless version of \eqref{02} reads
\be   
\epsilon K_{\tau \tau}=\frac{\partial }{\partial \zeta^2} \left[\hat{Q}(K^2)K+\hat{\nu}(K^2,\omega^2 K_{\tau}^2,\omega KK_{\tau})K_{\tau}\right] \label{02d}.
\en 

When $\hat{Q}=1$, $\nu=\nu_0$ choosing $\omega=\mu_0/\nu_0 $, (i.e. $\hat{\nu}=1$) from  \eqref{02d} we recover the classical  \emph{linear} Kelvin-Voigt material
\be
\epsilon K_{\tau \tau}=K_{\zeta \zeta}+K_{\tau \zeta \zeta}. \label{03}.
\en
  
If  in \eqref{03} we let $\epsilon \rightarrow 0$  we obtain the \emph{quasistatic} approximation.  This approximation is fundamental to model the usual creep and recovery experimental tests (Lakes, 2009). In Pucci and Saccomandi (2010) it has been shown that when:
\begin{itemize} 
\item the empirical inequalities are in force\footnote{In this case the generalized shear modulus, $Q$, is positive for any value of the amount of shear and the strain-energy density function associated with the elastic part of equation \eqref{02d} must be convex.} (Beatty, 1987);
\item the infinitesimal shear modulus $\mu_0$ is positive and finite; 
\item the viscosity function is such that 
$$
\lim_{K \rightarrow 0,K_t \rightarrow 0} \nu=\nu_0,
$$
where $\nu_0$ is a positive constant;
\end{itemize}
then the constitutive equations  \eqref{c1} are able to describe, at least qualitatively, the mechanical behavior observed in standard quasi-static experimental tests.

The quasi-static approximation is a \emph{singular perturbation} of the full dynamical equation \eqref{02d} and therefore is an approximation valid for large times. For this reason the global existence of solutions of \eqref{02d} is a main issue. The existence, uniqueness and regularity of solutions for initial-boundary value problems (IBVPs) associated to \eqref{02} (or its dimensionless version \eqref{02d}) have been studied by several authors. A state of art of the various results both for classical and weak solutions is provided in (Tvedt, 2008).  

The main problem with \eqref{02} is that the existence of smooth solutions global in time is not guaranteed when the viscosity function is nonlinear.  Indeed Tvedt  points out that:  
\begin{quote}
To my knowledge, all existence results for weak solutions of any kind to (1.1) in multiple space dimensions, including the Young-measure valued constructed in [6], require at least
$$
\left|S(p,q)\right|\leq C(1+\left|p\right|+\left|q\right|)
$$ 
\end{quote}
\begin{quote}
for some constant $C$. \emph{This condition is difficult to justify on physical grounds}.  
\end{quote}  
(Using the present notation $S=T_{13}$, $p=K$, $q=K_t$.).
 
It seems that the consequences of this negative result quoted by Tvedt are not full appreciated by workers  in nonlinear viscoelasticity. For example, if we have not a global existence result it is clear that we can consider a quasi-static approximation. Our show note is centered on the impact of this result in very simple problems of viscoelasticity. 

The plan of the paper is the following. In the next Section we derive some simple exact solutions for simple IBVPs with very smooth initial and boundary conditions of special cases of \eqref{02d} . For all these solutions there is a finite time $t^*$ associated with the blow-up or the extinction of the solution. In Section 3, we suggest that the \emph{catastrophic} problem associated with $t^*$ is located in the initial layer of the solution (i.e. for short times). By considering the case where the viscosity function depends only on the amount of shear we prove that the dynamics in the initial layer is governed by an equation equivalent to a nonlinear diffusion equation. The blow-up characteristics of such an equation have been studied into details in (Galaktionov and Vazquez, 1995) and the conditions derived there show the non-existence of global solutions. The last Section is devoted to concluding remarks.

\section{Some exact solutions}

We start by considering the problem associated with some viscous function that does not have an upper bound. To this end we consider  the following three nonlinear polynomial viscous functions  in turn
$$
\hat{\nu}_e(K^2)=1+\hat{\nu}_{e1}K^2, \quad  \hat{\nu}_v(K_\tau^2)=1+\hat{\nu}_{v1}K_\tau^2, \quad \hat{\nu}_m(KK_\tau)=1+\hat{\nu}_{m1}KK_\tau.
$$
For all these viscosity functions we have that $\lim_{K \rightarrow 0, K_\tau \rightarrow 0} \hat{\nu}=1>0$ but none of these functions are bounded from above. 

We start considering, for $\zeta \in [0,1]$,  the equation
\be \label{eq1}
\epsilon K_{\tau \tau}=K_{\zeta \zeta}+K_{\tau \zeta \zeta}+\hat{\nu}_{e1}\left(K^2K_\tau \right)_{\zeta \zeta},
\en
i.e. an equation corresponding to a material whose elastic part is neo-Hookean and with a super linear viscosity function in the amount of shear defined as $\hat{\nu}_e$.

This equation has an interesting and simple exact solution that may be computed as 
\be \label{ans}
K(\zeta, \tau)=\alpha(\tau)\zeta+\beta(\tau),
\en
where  
$$
\epsilon \frac{d \alpha}{d \tau}=2 \hat{\nu}_{e1} \alpha^3+c_1, \quad \epsilon \frac{d \beta}{d \tau}=2 \hat{\nu}_{e1} \alpha^2 \beta+c_2,
$$ 
(here $c_1$ and $c_2$ are two constants of integration).  If we consider $c_1=c_2=0$ and
$$
\beta(\tau)\equiv-\frac{1}{2}\alpha(\tau)\equiv\frac{1}{2}\sqrt{\frac{\epsilon}{\epsilon-\hat{\nu}_{e1}\tau}}
$$
we obtain 
\be
K(\zeta, \tau)=\frac{1}{2}\sqrt{\frac{\epsilon}{\epsilon-\hat{\nu}_{e1}\tau}}(\zeta-\frac{1}{2}) \rightarrow \hat{u}(\zeta, \tau)=\frac{1}{4}\sqrt{\frac{\epsilon}{\epsilon-\hat{\nu}_{e1}\tau}}(\zeta^2-\zeta) \label{04}.
\en
(Here $\hat{u}=u/L$).

The shearing motion in \eqref{04} satisfies the following set of smooth initial and boundary conditions 
$$
\hat{u}(0,\tau)=\hat{u}(1,\tau)=0, \, \hat{u}(\zeta,0)=(\zeta^2-\zeta)/4, \, \hat{u}_{\tau}(\zeta,0)=\hat{\nu}_1(\zeta^2-\zeta)/(8 \epsilon).
$$
In this way we have provided a simple example of non-existence (because of finite time blow-up) of the solution for a very regular IVBPs  for equation \eqref{eq1} in $[0,1]$. 

If now we consider the equation
\be \label{eq2}
\epsilon K_{\tau \tau}=K_{\zeta \zeta}+K_{\tau \zeta \zeta}+\hat{\nu}_{v1}\left(K_\tau^3 \right)_{\zeta \zeta},
\en
a similar solution may be obtained using the ansatz in \eqref{ans}  for the amount of shear. The exact solution is now
\be
\hat{u}(\zeta, \tau)=\frac{1}{2}\sqrt{\frac{12 \hat{\nu}_{v1}-\epsilon\tau}{3\hat{\nu}_{v1}}}(\zeta^2-\zeta) \label{04bis}.
\en
This solution may be used to solve the boundary-initial value problem
$$
\hat{u}(0,\tau)=\hat{u}(1,\tau)=0, \, \hat{u}(\zeta,0)=\zeta^2-\zeta, \, \hat{u}_{\tau}(\zeta,0)=-\frac{\epsilon}{24\hat{\nu}_{{v1}}}(\zeta^2-\zeta).
$$
Here we record that the reason why the existence of the solution fails  in finite time is the extinction phenomena.

The last equation we consider is
\be \label{eq3}
\epsilon K_{\tau \tau}=K_{\zeta \zeta}+K_{\tau \zeta \zeta}+\hat{\nu}_{m1}\left(KK_\tau^2 \right)_{\zeta \zeta}.
\en
Here the mathematics is more involved. Let us consider that the shearing motion is in the form
$$
\hat{u}(\zeta,\tau)=\alpha(\tau)(\zeta^2-\zeta).
$$
Then the time dependent function $\alpha$ is determined by the ordinary differential equation
$$
\frac{\epsilon}{\hat{\nu}_{m1}} \alpha''=24\alpha \alpha'^2.
$$
The solution of this second order ordinary differential equation is given by
$$
\text{Erf}\left(\tilde{\alpha}\right)=c_1 \tau+c_2,
$$ where 
$$
\tilde{\alpha}=2\sqrt{\frac{3\hat{\nu}_{m1}}{\epsilon}}\alpha,
$$
Erf is the standard error function and $c_1, \,c_2$ two integration constants.
This exact solution solves the following IBVP (in the normalized variables) 
$$
\tilde{u}(0,\tau)=\tilde{u}(1,\tau)=0, \, \tilde{u}(\zeta,0)=\delta(\zeta^2-\zeta),  
$$
with $0<\delta<1$ and $\tilde{u}_{\tau}(\zeta,0)=\mu(\zeta^2-\zeta)$ where $\mu$ is an ad hoc constant determined by $\alpha'$. From
\be \label{04tris}
\text{Erf}\left(\tilde{\alpha}\right)-\text{Erf}\left(\delta\right)=c_1 \tau,
\en
it is clear that this solution blows up in a finite time.  

To summarize, the following remarks apply to all the three examples we have proposed. First of all in the case $\epsilon \rightarrow 0$ all the exact solutions we propose degenerate and are of no interest. Second, for all three solutions there exists a critical time $\tau^*$ such that for $\tau>\tau^*$ the solution does not exists anymore. For the solution in \eqref{04} and \eqref{04tris} we record a blow-up phenomenon and for the solution \eqref{04bis} we have an extinction phenomenon. 

Therefore we have provided explicit examples of the lack of global existence  for equations with nonlinear viscous function and a neo-Hookean elastic term. For all these solutions  the inertia term (i.e. $\epsilon \neq 0$) is  crucial in delineating why the global existence fails.  

%%%%%%%%%%%%%%%%%%%%%%%%%%%%%%%%%%%%%%%%%%%%%%

\section{The initial layer}
 
%%%%%%%%%%%%%%%%%%%%%%%%%%%%%%%%%%%%%%%%%%%%%%
We point out once again that in the quasi-static limit all the models we have considered in the previous Section are "good" models. This fact suggests  that because the quasi-static limit is the result of a singular perturbation procedure, the failure of well-posedness must occur for short times or, using the terminology of singular perturbations in the \emph{initial layer}.  

When we consider a general viscosity function defined as $\hat{\nu}=\hat{\nu}(K^2)$ and we introduce the \emph{stretched} time $\tilde{\tau}=\tau/\epsilon$, it is possible to obtain easily, for any regular viscosity function, the equation governing the dynamics in the initial layer. Restricting the attention to this class of materials it is then possible to rewrite \eqref{02d} as
\be   
K_{\tilde{\tau} \tilde{\tau}}=\frac{\partial }{\partial \zeta^2} \left[\epsilon \hat{Q}(K^2)K+\hat{\nu}(K^2)K_{\tilde{\tau}}\right] \label{02e},
\en 
and when $\epsilon \rightarrow 0$ we derive the equation
\be   
K_{\tilde{\tau} \tilde{\tau}}=\frac{\partial^3 \Pi }{\partial \zeta^2 \partial \tilde{\tau}} \label{02f}.
\en 
where $\Pi=\int \hat{\nu}(K^2)dK$. 

By a simple integration the above equation is rewritten as
\be   
K_{\tilde{\tau}}=\frac{\partial^2 \Pi }{\partial \zeta^2}+\varphi(\zeta) \label{02g}.
\en 
Equation \eqref{02g} is a nonlinear diffusion equation. This equation has been considered by many authors and the necessary and sufficient conditions for the complete blow-up and extinction of this equation have been considered in (Galaktionov and Vazquez, 1995). It is interesting to note that for equation \eqref{02g} we have blow-up of the solution in finite time if the integral
$$
\int_{1}^{\infty}\frac{\Pi'(s)}{s}ds,
$$
does not converge. (Here a dash denotes  the derivative with respect the amount of shear). This is exactly the case of the examples in the previous Section. 

It is clear that there are several choices of the viscosity functions for which the above integral  may converge, but the usual higher order polynomial models that are used in the standard fitting procedures do blow up. We remark that a model with a constant viscosity, in the initial layer, is associated with a linear heat equation this explain why it  is possible to find a unique global solution for such a model in the class \eqref{c1}. 

We remark that via a multiple scale methods is possible to deduce the equation in the initial layer for more general viscosity functions, but the case we have just examined is sufficient to point out the problems inherent with the constitutive equation \eqref{c1}  when the sub linear growth condition pointed out by Tvedt (2009) is not satisfied.

%%%%%%%%%%%%%%%%%%%%%%%%%%%%  

\section{Concluding remarks}

%%%%%%%%%%%%%%%%%%%%%%%%%%%%
If we believe that materials of differential type have a "good" quasi-static behavior, the fact that a global solution for the full dynamical equations does not exist is a major problem. Indeed, if a solution blows up, or is associated with an extinction phenomena, in finite time it is not possible to consider the asymptotic limit for $t\rightarrow \infty$ and therefore it is not possible to consider the quasi-static approximation. 

Clearly such situation may be arranged considering a sort of "cut-off" for the usual polynomial constitutive equations. For example we may be consider a constitutive equation such as 
\[ \hat{\nu}(K^2) = \left\{ \begin{array}{ll}
         1+\nu_{e1}K^2 & \mbox{if $K^2 \leq K^*$};\\
        \nu_{e} & \mbox{if $K^2 >  K^*$}.\end{array} \right. \] 

We cannot be satisfied with this approach for two reasons.  First of all we introduce some oddities in the quasi-static regime as it is possible to check by analyzing the corresponding creep problem that loose regularity. Moreover, we do not solve the "unphysical" situation generated by the fact: greater viscous effects lead to greater problems with the regularity of the solution.  
 
This situation seems to suggest that there is something missing in the "dynamics" of the nonlinear viscoelastic models we are considering. It seems that the model \eqref{c1} must be refined taking into account phenomena that at first sight seemed to be irrelevant but indeed are fundamental. These phenomena seems to be of the same order of the inertia terms and therefore disappear when $\epsilon \rightarrow 0$. 

A possibility is to consider  the presence of objective derivatives of the stretching tensor $\vec{D}$ i.e. higher order models of differential type where the constitutive equations not only depends on the stretching tensor $\vec{D}$ but for example also on the subsequent Rivlin-Ericksen tensors (Truesdell and Noll, 1965). These tensors may introduce some additional terms that regularize the equations and allows to avoid the problems presented here.  This direction is outlined in the context of non dissipative materials in (Destrade and Saccomandi, 2006).

%%%%%%%%%%%%%%%%%%%%%%

\section*{Acknowledgements}

%%%%%%%%%%%%%%%%%%%%%%

The research is supported by PRIN-2009 project \emph{Matematica e meccanica dei sistemi biologici e dei tessuti molli}.

GS thanks Michel Destrade and Jeremiah Murphy for their remarks on a previous draft of this paper. 

This paper is dedicated with great admiration to Professor Robin Knops for his 80-th birthday. 
 
%%%%%%%%%%%%%%%%%%%%%%%%%%%%  

%%%%%%%%%%%%%%%%%%%%%%

\end{document}